\documentclass[aps,showkeys]{revtex4}
\usepackage{amsfonts}
\usepackage{amsmath}
\usepackage{amssymb}
\usepackage{graphicx}

\providecommand{\U}[1]{\protect\rule{.1in}{.1in}}

\input{tcilatex}
\begin{document}

\title{Fermion interactions, cosmological constant and space-time
dimensionality in a unified approach based on affine geometry}
\author{Salvatore Capozziello$^{1,2}$, Diego Julio Cirilo-Lombardo$^3$,
Alexander E. Dorokhov$^3$}
\affiliation{$^1$Dipartimento di Fisica, Universita' di Napoli \textquotedblleft Federico
II\textquotedblright, and $^2$INFN Sez. di Napoli, Complesso Universitario
di Monte S. Angelo, Via Cinthia, Napoli I-80126, Italy}
\affiliation{$^3$Bogoliubov Laboratory of Theoretical Physics,Joint Institute for Nuclear
Research 141980, Dubna(Moscow Region), Russian Federation}
\pacs{PACS number}

\begin{abstract}
One of the main features of unified models, based on affine geometries, is
that all possible interactions and fields naturally arise under the same
standard. Here, we consider, from the effective Lagrangian of the theory,
the torsion induced 4-fermion interaction. In particular, how this
interaction affects the cosmological term, supposing that a condensation
occurs for quark fields during the quark-gluon/hadron phase transition in
the early universe. We explicitly show that there is no parity-violating
pseudo-scalar density, dual to the curvature tensor (Holst term) and the
spinor-bilinear scalar density has no mixed couplings of A-V form. On the
other hand, the space-time dimensionality cannot be constrained from
multidimensional phenomenological models admitting torsion.
\end{abstract}

\maketitle





\section{Introduction}

As it is well known, the Standard Model (SM) of particle physics has proven
to be a very successful theory, but still presents some shortcomings. Among
these, one is the Hierarchy Problem, which the current literature claims
that might be solved by considering extra dimensions. In higher dimensional
spacetimes, the gravitational theory by Einstein can be generalized in
several ways: the Kaluza-Klein approach (or slight modifications of it) is
currently used by a great percentage of the higher dimensional models
treating the hierarchy trouble (see, for example, \cite{basini}).

Another important generalization of the General Relativity (GR) is the
gravity with torsion that considers the possibility of a general
non-symmetric connection. This point of view is also assumed in several
Extended Theories of Gravity like $f(R)$-gravity \cite{rept,cianci,fabbri}
that have recently gained interest for dark energy and dark matter issues 
\cite{francaviglia, sergei}. The first direct geometrical generalization in
this direction is due to Cartan \cite{cartan}. The approach taken into
account by theoretical physicists in the last decades is that, assuming as
the starting point Cartan's generalization of GR, torsion can be coupled to
fermion matter in a straightforward way. The trick in such approaches is,
since the field equation for the spin connection is a constraint related to
the contortion, it can be used to get rid of the torsion in the original
action. Then the torsion field is a non-dynamical one and the fermionic
matter is added by hand. Consequently, the new "artificial" action contains
standard GR and matter fields with an additional contact 4-fermion
interaction \citep{1} where the Dirac equation for the fermions is not
derived from the geometrical structure of space-time.

Because of the effective 4-fermion interaction term has a coupling constant
proportional to Newton's gravitational constant, at first approximation,
this interaction is highly suppressed. Nevertheless, it is currently claimed
that extra dimensions could explain the hierarchy problem, and thus the
(higher dimensional) fundamental gravity scale might be roughly $%
M^{\ast}\sim O(1)$ TeV \citep{basini, 2,3,4,5}. The limits to the size of
extra dimensions have been set up by direct searches for quantum black holes 
\cite{6} and the exchange of virtual gravitons on di-lepton events \cite{7}.
On the other hand, the ATLAS collaboration has presented experimental limits
for the coupling constant of 4-fermion contact interaction \cite{7,8}. These
results are currently used for imposing bounds on the value of the
fundamental gravity scale, $M^{\ast}$, and, by extension, in order to find
limits on the dimensionality of the space-time. However, as we will show
here, these claims could have shortcomings from the theoretical and
phenomenological viewpoints.

On the other hand, there exists the cosmological constant problem, that is
repeatedly faced not only from the Quantum Field Theory (QFT) point of view
but also from the Quantum Gravity and early cosmology viewpoints. There are
many mechanisms and scenarios trying to explain consistently the problem,
some of them against the physical intuition. For example, recently Brodsky
et al. have argued that quark and gluon condensates are spatially restricted
to the interiors of hadrons and do not extend throughout all of space \cite%
{9,10}. Such argument seems to have some problem. Consequently, alternative
possibilities need to be studied and developed. The other problem that
arises here is how to generate the 4-fermion interaction from first
principles and how to get its contributions to the cosmological constant. An
approach where cosmological constant comes out from fermion condensations is
discussed in \cite{capolupo}.

The question that immediately arises is if there exist other mechanisms to
explain faithfully the 4-fermion interaction without the drawbacks inherent
to the standard Einstein-Cartan theory. Some affirmative answers are
possible to this question, as we will discuss below.

Our argument is based on a gravity theory where a pure affine geometry is
adopted with the gravitational Lagrangian given by 
\begin{equation}
L_{g}=\sqrt{det(\mathcal{R}_{\ \mu }^{a}\mathcal{R}_{a\nu })}  \tag{1}
\end{equation}%
where the specific Ricci curvature tensor is determined by 
\begin{equation}
\mathcal{R}_{\ \mu }^{a}=\lambda \left( e_{\ \mu }^{a}+f_{\ \mu }^{a}\right)
+R_{\ \mu }^{a},\qquad ,  \tag{2}  \label{R}
\end{equation}%
that corresponds to the breaking of the $SU(2,2)$ symmetry of a group
manifold in higher dimensions with original Riemann curvature: $\mathcal{R}%
_{\mu \nu }^{AB}=\partial _{\mu }\omega _{\nu }^{AB}-\partial _{\nu }\omega
_{\mu }^{AB}+\omega _{\mu }^{AC}\omega _{\nu C}^{\ \ B}-\omega _{\nu
}^{AC}\omega _{\mu C}^{\ \ B}$ (see \cite{11},\cite{11a} for details) to the 
$SO(2,2)$ group. The absolute value of the determinant in (1) is assumed.
However, imposing (anti) self-duality conditions over the generalized
curvature $\mathcal{R}$, an Euclidean condition is obtained In eq. (\ref{R}%
), $e_{\ \mu }^{a}$ is the tetrad field 
\begin{equation}
g_{\mu \nu }\equiv e_{\mu }^{a}e_{\nu a}\text{ \ , \ \ }\eta _{ab}\equiv
e_{a}^{\nu }e_{\nu b}\text{\ \ }  \tag{5}
\end{equation}%
and $f_{\ \mu }^{a}$ is antisymmetric with respect to the index permutation%
\begin{equation}
e_{a\mu }f_{\ \nu }^{a}=f_{\mu \nu }=-f_{\nu \mu },  \tag{3}
\end{equation}%
which is associated with a central tensorial part of the original $SU(2,2)$
group. Both $e_{\ \mu }^{a}$ and $f_{\ \mu }^{a}$ can be taken as
fundamental fields from which the Palatini variational principle is applied $%
R_{\mu \nu }$ is the Ricci curvature tensor in a manifold with torsion, $M$
(e.g. $U_{4}$) and $\lambda =(1-d)$ with $d$ being the spacetime dimension.
Notice, that the Ricci tensor has symmetric and antisymmetric parts
corresponding to the Christoffel and torsion contributions to the
connection. Here and below, we consider Greek letters $\mu ,\nu $ as
coordinates indices and Latin letters $a,b$ as tetrad indices. With this
formalism, we have $M_{\mu }^{a}\equiv e^{a\nu }M_{\nu \mu }$. 
By using eq. (\ref{R}), the Lagrangian becomes 
\begin{equation}
L_{g}=\sqrt{\det \left[ \lambda ^{2}\left( g_{\mu \nu }+f_{\ \mu
}^{a}f_{a\nu }\right) +2\lambda R_{\left( \mu \nu \right) }+2\lambda f_{\
\mu }^{a}R_{[a\nu ]}+R_{\ \mu }^{a}R_{a\nu }\right] },  \tag{4}
\end{equation}%
where the Ricci tensor can be split in its symmetric and anti-symmetric
part: $R_{\mu \nu }=R_{(\mu \nu )}+R_{[\mu \nu ]}$ (see reference \cite%
{11,11a,11b, 12,13, 14} for details). The basis of the considered approach
is a hypercomplex construction of the (metric compatible) space-time
manifold $M$ \cite{11, 11a,11b, 12, 13, 14}, where for each point of $M$
there exists a local affine space $A.$ The connection over $A,$ $\ 
\widetilde{\Gamma }$, defines a generalized affine connection $\Gamma $ on $%
M $ specified by $\nabla $ and $K$, where $K$ is an invertible $\left(
1,1\right) $ tensor over $M.$ Connection is compatible and rectilinear, i.e. 
\begin{equation}
\nabla _{\mu }K_{\rho \sigma }=K_{\rho \alpha }T_{\mu \sigma }^{\alpha
},\;\;\;\;\;\text{ }\nabla _{\mu }g_{\mu \nu }=0,  \tag{6}
\end{equation}%
where $T_{\mu \sigma }^{\alpha }$ is the torsion tensor and $g_{\mu \nu }$
is the metric tensor preserved under parallel transport. This compatibility
condition ensures that the affine connection $\widetilde{\Gamma }$ maps
auto-parallel curves of $\Gamma $ on $M$ in straight lines over the affine
space $A$ (locally). The first equation is the condition determining the
connection $\Gamma $ in terms of the fundamental tensor $K$. 

As it is well known, the Palatini variational principle determines the
connection required for the space-time symmetry as well as the field
equations. From here we assume a four dimrnsional spacetime. Consequently
and by construction, the action $(1)$ yields the $G$-invariant conditions
(namely, the intersection of the 4-dimensional Lorentz group $L_{4},$ the
symplectic $Sp\left( 4\right) $ and the almost complex group $K\left(
4\right) )$, without prior assumptions. As a consequence, the gravitational,
Dirac and Maxwell equations arise from the Lagrangian $L_{g}$ as a causally
connected closed system. The self-consistency is given by 
\begin{equation}
f_{\mu \nu }\equiv \frac{1}{2}\varepsilon _{\mu \nu \rho \sigma }\varphi
^{\rho \sigma }=\ast \varphi _{\mu \nu }  \tag{7}
\end{equation}%
where $\varphi _{\nu \lambda }$ is related to the torsion by ${\displaystyle%
\frac{1}{6}\left( \partial _{\mu }\varphi _{\nu \lambda }+\partial _{\nu
}\varphi _{\lambda \mu }+\partial _{\lambda }\varphi _{\mu \nu }\right)
=T_{\ \nu \mu }^{\rho }\varphi _{\rho \lambda }}$ and $f_{\mu \nu }$ plays
the role of electromagnetic field.

As it was shown in \cite{11,11a,11b,12,13,14} for this model of gravity (see
Ref.\cite{15}, for astrophysical neutrino applications), the Dirac equation
is derived from the same space-time manifold and acquires a coupling
modification of the form 
\begin{equation}
\gamma^{\alpha}j\left(\frac{1-d}{d}\right)\gamma_{5}h_{\alpha},  \tag{8}
\end{equation}
where $h_{\alpha}=\varepsilon_{\alpha}^{\text{ }\nu\rho\sigma}T_{\text{ }%
\nu\rho\sigma}$ is the torsion vector defined by the duality operation in
4-dimensions and $\ j$ is a parameter of pure geometrical nature. Here, the
torsion described by $h_{\alpha}$ is a dynamical field and the theory is
Lorentz invariant by construction. This dynamical torsion vector is
responsible for generation of the 4-fermion interaction, as it will be shown
below.

The aim of this work is twofold: first, we discuss the possibility to
explain the nature, magnitude, bounds and contributions to the cosmological
constant due to the 4-fermion interaction from the point of view of unified
theories with torsion based on affine geometries. Secondly, we compare our
approach with other attempts coming from the context of Riemann-Cartan
theory.

The layout of the paper is the following. In Sec.II, we discuss the fermion
interaction and vector torsion in view of the Hodge - de Rham decomposition.
Sec.III is devoted to the derivation of the gravitational field and Dirac
equations, while in Sec.IV, we discuss the fermionic structure. The
effective action and the 4-fermionic interaction, together with the
energy-momentum tensor, are considered in Sec. V. Conclusions are drawn in
Sec.VI.

\section{Generalized Hodge-de Rham decomposition, the vector torsion $h$ and
the fermion interaction}

As pointed out in references \cite{11, 11a,11b, 12,13,14}, the torsion
vector $h=h_{\alpha}dx^{\alpha}$ (the 4-dimensional dual of the torsion
field $T_{\beta\gamma\delta}$) plays multiple roles and can be constrained
in several different physical situations. Mathematically, it is defined by
the Hodge-de Rham decomposition given by the \textbf{4-dimensional Helmholtz
theorem} which states:

\textit{If $h=h_{\alpha }dx^{\alpha }$ $\notin F^{\prime }\left( M\right) $%
(set of derivative of functions on M) is a 1-form on $M$, then there exist a
zero-form $\Omega $, a 2-form $\alpha =A_{\left[ \mu \nu \right] }dx^{\mu
}\wedge dx^{\nu }$ and a harmonic 1-form $q=q_{\alpha }dx^{\alpha }$ on $M$
that}%
\begin{equation*}
h=d\Omega +\delta \alpha +q\rightarrow h_{\alpha }=\nabla _{\alpha }\Omega
+\varepsilon _{\alpha }^{\beta \gamma \delta }\nabla _{\beta }A_{\gamma
\delta }+q_{\alpha }\,.
\end{equation*}%
%
%
%
%
%
%
%
%
Notice that even if $q_{\alpha }$ is not harmonic, and assuming that $%
q_{\alpha }=$ $\left( P_{\alpha }-eA_{\alpha }\right) $ is a vector, an
axial vector can be added such that the above expression takes the form%
\begin{align}
h_{\alpha }& =\nabla _{\alpha }\Omega +\varepsilon _{\alpha }^{\beta \gamma
\delta }\nabla _{\beta }A_{\gamma \delta }+\varepsilon _{\alpha }^{\beta
\gamma \delta }M_{\beta \gamma \delta }+\left( P_{\alpha }-eA_{\alpha
}\right)  \tag{10} \\
& =\nabla _{\alpha }\Omega +\varepsilon _{\alpha }^{\beta \gamma \delta
}\nabla _{\beta }A_{\gamma \delta }+b_{\alpha }+\left( P_{\alpha
}-eA_{\alpha }\right) \,,  \notag
\end{align}%
where $M_{\beta \gamma \delta }$ is a completely antisymmetric tensor. In
such a way, $\varepsilon _{\alpha }^{\beta \gamma \delta }M_{\beta \gamma
\delta }$ $\equiv b_{\alpha }$ (axial vector).

One can immediately see that, due to the theorem given above, one of the
roles of $h_{\alpha}$ is precisely to be a generalized energy-momentum
vector, avoiding the addition "by hand" of a matter Lagrangian in the action
(4). As it is well known, the addition of the matter Lagrangian leads, in
general, to non-minimally coupled terms into the equations of motion of the
physical fields. Consequently, avoiding the addition of energy-momentum
tensor, the fields and they interactions are effectively restricted thanks
to the same geometrical structure in the space-time itself.

\section{Gravitational field and Dirac equations}

It is possible to show \cite{11,11a,11b,12,13,14} that to derive the Dirac
equation, one needs, as starting point, the symmetric part of the
gravitational field equations derived from $\delta _{g}L_{g}=0$, that is 
\begin{align}
\overset{\circ }{R}_{\mu \nu }& =-2\lambda g_{\mu \nu }+T_{\mu \rho }^{\ \ \
\alpha }T_{\alpha \nu }^{\ \ \ \rho }  \tag{11} \\
& =-2\lambda g_{\mu \nu }-2w\left( g_{\mu \nu }h_{\alpha }h^{\alpha }-h_{\mu
}^{\ \ \ }h_{\nu }^{\ \ \ }\right) =-2\lambda g_{\mu \nu }-2\left( g_{\mu
\nu }\Pi _{\alpha }\Pi ^{\alpha }-\Pi _{\mu }\Pi _{\nu }\right)  \tag{12}
\end{align}%
Here we use the obvious duality relation between $\ T$ and $h$ and define
the generalized momentum vector as: $\sqrt{w}h_{\mu }^{\ \ \ }=\Pi _{\mu }$
with $w$ some arbitrary constant that will be conveniently fixed. Then, a
mass-like shell condition is immediately obtained 
\begin{equation}
\Pi ^{2}=m^{2}\Rightarrow m=\pm \sqrt{\frac{\overset{\circ }{R}}{2(1-d)}+d}%
\,.  \tag{13}
\end{equation}%
where the definition mass (where the mass-like shell hold true) is connected
with the spacetime structure, due the unified character of the theory.
Notice that there exists a link between the dimension of the spacetime and
the scalar "Einstenian" curvature $\overset{\circ }{R}$. Moreover, the
curvature and the mass are constrained to take definite values in order that 
$d\in $ $\mathbb{N}$, the natural number characteristic of the dimension. On
the other hand, knowing that $\lambda =1-d$ and assuming that the parameter $%
m$ $\in \mathbb{R}$ , the limiting condition on the physical values for the
mass is $\frac{\overset{\circ }{R}}{2(1-d)}+d\geqslant 0$.

Admitting $\Pi _{\mu }\rightarrow \widehat{P}_{\mu }-e\widehat{A}_{\mu
}+\gamma ^{5}b_{\mu }$ (with $b_{\mu }\equiv \epsilon _{\mu }^{\text{ }\nu
\rho \sigma }M_{\nu \rho \sigma }$ an axial vector) together with the
quantum condition where the classical equation is converted to an operator: $%
\Pi _{\mu }\rightarrow \widehat{\Pi }_{\mu }$, we have%
\begin{equation}
\left\{ \left[ \gamma ^{\mu }\left( \widehat{P}_{\mu }-e\widehat{A}_{\mu
}+c_{1}\gamma ^{5}b_{\mu }\right) +m\right] \left[ \gamma ^{\nu }\left( 
\widehat{P}_{\nu }-e\widehat{A}_{\nu }+c_{1}\gamma ^{5}b_{\nu }\right) -m%
\right] \right\} \Psi =0\,,  \tag{14}
\end{equation}%
(where $\Psi =\mathbf{u}+i\mathbf{v}$ is a complex function) which leads to
the Dirac equation 
\begin{equation}
\left[ \gamma ^{\mu }\left( \widehat{P}_{\mu }-e\widehat{A}_{\mu
}+c_{1}\gamma ^{5}\widehat{b}_{\mu }\right) -m\right] \Psi =0\,,  \tag{15}
\end{equation}%
with $m$ given by (13). Notice that this condition, in the Dirac case, is
not obtained only passing from classical variables to quantum operators, but
in the case that the action does not contain explicitly $\widehat{A}_{\mu }$%
, $h_{\mu }$ remains without specification due the gauge freedom in the
momentum. Notice that the unified character of the theory make the number of
equations and the field transformations above self consistent, as will be
clear in Section V( see the effective action). From the second order version
of (14), it is not difficult to show that for $u^{\lambda }$ (remind $\Psi =%
\mathbf{u}+i\mathbf{v):}$ 
\begin{gather}
\left\{ \left( \widehat{P}_{\mu }-e\widehat{A}_{\mu }+c_{1}\gamma ^{5}%
\widehat{b}_{\mu }\right) ^{2}-m^{2}-\frac{1}{2}\sigma ^{\mu \nu }\left[ e%
\underset{\equiv F_{\mu \nu }}{\underbrace{\left( \nabla _{\mu }\widehat{A}%
_{\nu }-\nabla _{\nu }\widehat{A}_{\mu }\right) }}-c_{1}\gamma ^{5}\underset{%
\equiv S_{\mu \nu }}{\underbrace{\left( \nabla _{\mu }\widehat{b}_{\nu
}-\nabla _{\nu }\widehat{b}_{\mu }\right) }}\right] \right\} u^{\lambda } 
\tag{16} \\
+\frac{1}{2}\sigma ^{\mu \nu }R_{\rho \left[ \mu \nu \right] }^{\lambda
}u^{\rho }-\frac{1}{2}e\sigma ^{\mu \nu }\left( \widehat{A}_{\mu }\widehat{P}%
_{\nu }-\widehat{A}_{\nu }\widehat{P}_{\mu }\right) u^{\lambda }=0  \notag
\end{gather}%
(the same, obviously, for $v^{\lambda }).$It is interesting to see that eq.
(16) differs from that obtained by the standard expression derived in \cite%
{16} due to the appearance of the last two terms: the term involving the
curvature tensor is due to the spin interaction with the gravitational field
(due to the torsion term in $R_{\rho \left[ \mu \nu \right] }^{\lambda })$
and the last term is the spin interaction with the electromagnetic and
mechanical momenta. The important point here is that the spin-gravity
interaction term is derived since the spinors are represented as space-time
vectors whose covariant derivatives are defined in terms of the G-(affine)
connection (see also \cite{cosimo} for the classification of torsion
tensor). \ Other important point to remark is that, in order for Dirac
equation to be global (or being covariant with respect to spin
transformations if working locally) global topological conditions on M are
needed.Through this work, we are working locally and global issues can be
taken into account as in standard references (e.g. \cite{25} or references
quoted therein and \cite{26},\cite{27} for the relation between spin
structures and Dirac equations).

\section{Fermionic structure, electromagnetic field and anomalous
gyromagnetic factor}

If we introduce an expression corresponding to the antisymmetric part of the
gravitational field, namely $\nabla_{\alpha}T^{\alpha\beta\gamma}=-2\lambda
f^{\beta\gamma}$, in (16) then

\begin{gather}
\left[ \left( \widehat{P}_{\mu}-e\widehat{A}_{\mu}+c_{1}\gamma^{5}\widehat{b}%
_{\mu}\right) ^{2}-m^{2}-\frac{1}{2}\sigma^{\mu\nu}\left(
eF_{\mu\nu}-c_{1}\gamma^{5}S_{\mu\nu}\right) \right] u^{\lambda}-\frac{%
\lambda}{d}\frac{1}{2}\sigma^{\mu\nu}f_{\left[ \mu\nu\right] }u^{\lambda} 
\tag{17} \\
-\frac{1}{2}e\sigma^{\mu\nu}\left( \widehat{A}_{\mu}\widehat{P}_{\nu }-%
\widehat{A}_{\nu}\widehat{P}_{\mu}\right) u^{\lambda}=0  \notag
\end{gather}
as a consequence, we have 
\begin{equation}
\left[ \left( \widehat{P}_{\mu}-e\widehat{A}_{\mu}+c_{1}\gamma^{5}\widehat{b}%
_{\mu}\right) ^{2}-m^{2}-\frac{1}{2}\sigma^{\mu\nu}\left( eF_{\mu\nu}%
\underset{anomalous\text{ }term}{\underbrace{-c_{1}\gamma^{5}S_{\mu\nu}+%
\frac{\lambda}{d}f_{\mu\nu}+e\left( \widehat{A}_{\mu}\widehat {P}_{\nu}-%
\widehat{A}_{\nu}\widehat{P}_{\mu}\right) }}\right) \right] u^{\lambda}=0 
\tag{18}
\end{equation}
where clearly appear the contributions to the (g-2) factor due the axial
vector $\widehat{b}_{\mu}$ and the geometry through the commutation relation
between the covariant derivatives $\nabla$. We can go ahead and see that if $%
\omega_{2}F_{\mu\nu}=S_{\mu\nu}$ and $\omega_{1}F_{\mu\nu}=\sigma_{\mu\nu
}^{\prime}$ the last expression assumes the suggestive form%
\begin{gather}
\left\{ \left( \widehat{P}_{\mu}-e\widehat{A}_{\mu}+c_{1}\gamma^{5}\widehat{b%
}_{\mu}\right) ^{2}-m^{2}-\frac{1}{2}\sigma^{\mu\nu}\left[ \left(
e-c_{1}\omega_{2}\gamma^{5}-\omega_{1}\frac{\lambda}{d}\right) F_{\mu\nu
}+\omega_{1}\frac{\lambda}{d}\sigma_{\mu\nu}\right] \right\}u^{\lambda }- 
\tag{19} \\
-\frac{e}{2}\sigma^{\mu\nu}\left( \widehat{A}_{\mu}\widehat{P}_{\nu}-%
\widehat{A}_{\nu}\widehat{P}_{\mu}\right) u^{\lambda}=0  \notag
\end{gather}
with the result that the gyromagnetic factor results modified accordingly,
and a 4-fermion coupling is introduced constructively, thanks to $f_{\mu\nu}$%
. Although the anomalous term is clearly determined from the above
equations, it is extremely useful in order to compare the present scheme to
other theoretical approaches. 

With these considerations in mind, it is importantl to derive the anomalous
momentum for the electron. Specifically, from the last expression, one gets
the correction to the lepton anomalous momentum in the form%
\begin{equation*}
\Delta a_{e}=-\frac{\omega _{1}}{e}\frac{\lambda }{d}\equiv \frac{\omega _{1}%
}{e}\left( 1-\frac{1}{d}\right) .
\end{equation*}%
The experimental precision in measurement of this quantity is \cite{17} 
\begin{equation*}
\Delta a_{e}^{\exp }=0.28\times 10^{-12}
\end{equation*}%
and then the upper bound for the universal geometric parameter $\omega _{1}$
is%
\begin{equation*}
\omega _{1}<e\left( \frac{d-1}{d}\right) 0.28\times 10^{-12}
\end{equation*}%
then, in 4-dimensions, we have $\omega _{1}<\frac{3}{4}e$ $0.28\times
10^{-12}$. This result is useful in order to give constraints to the theory.


Another important consideration is related to the anomalous magnetic
momentum. As it is well known from the quantum point of view, in the
lowest-order diagram, the anomalous magnetic momentum term is given by 
\begin{equation*}
\Delta \Gamma _{\mu }\left( p,p^{\prime }=p+q\right) =-2iA^{2}\int \frac{%
d^{4}k}{\left( 2\pi \right) ^{4}}\Gamma _{a}S\left( k^{\prime }=k+q\right)
\gamma _{\mu }S\left( k\right) \Gamma _{a}
\end{equation*}%
where%
\begin{equation*}
S\left( k\right) =\frac{1}{\widehat{k}-M\left( k\right) }
\end{equation*}%
is the formal propagator and $\Gamma _{a}=\left\{ I,\gamma _{\nu },\gamma
_{\nu }\gamma _{5}\right\} $ for $a=S,P,V,A,T$ (scalar, pseudoscalar,
vector, axial vector and tensor bilinears). The Fierz transformation for the
integrand, necessary to reduce and rewrite the matrix quantities in a
convenient computational form, has a general form as%
\begin{align*}
& \Gamma _{a}\left( \widehat{k^{\prime }}-M\left( k^{\prime }\right) \right)
\gamma _{\mu }\left( \widehat{k}-M\left( k\right) \right) \Gamma _{a} \\
& =\sum_{\alpha =s,p,v,a,t}C_{\alpha }\mathrm{Tr}\left[ \left( \widehat{%
k^{\prime }}+M\left( k^{\prime }\right) \right) \gamma _{\mu }\left( 
\widehat{k}+M\left( k\right) \right) \Omega _{\alpha }\right] \Omega
_{\alpha },
\end{align*}%
where $C_{\alpha }$ are the coefficients of the Fierz transformation. This
makes the link between experimental data and theory, as masses, phase space
and matrix elements from cross sections. See also \cite{20} for details.

\section{Effective action, $\Theta$ term and 4-fermion interaction}

Now, let us analyze the theory from a different point of view. In some
gravitational models, a link between torsion and CP violating terms
certainly appears. An illustrative example is given by Ashtekar that has
rewritten Einstein's theory, in its Hamiltonian formulation, as a set of
differential equations obeyed by an SO(3) connection and its canonically
conjugate momenta corresponding to the SO(3) gauge \cite{ashtekar}.
Bengtsson and Peldan \cite{18} have shown that if one performs a particular
canonical transformation involving Ashtekar's variables and the
corresponding SO(3) gauge fields, the expression for the Hamiltonian
constraint changes when other constraints remain unaffected. This
corresponds precisely to the addition of a "CP-violating"-term to the
corresponding Lagrangian. Mullick and Bandyopadhyay have shown that this
CP-violating-term is responsible for nonzero torsion \cite{19}. This $\theta 
$-term effectively corresponds to the chiral anomaly when a fermion chiral
current interacts with a gauge field. Here, the contrary statement is found
from first principles: the theory with torsion leads, at effective level, a $%
\theta $-term directly related to the space-time dimensions through the
"cosmological" constant $\lambda =1-d$ as we will see soon.

\subsection{\textbf{Deriving the effective Lagrangian}}

As in the case of massive vector particle with spin 1, let us derive the
effective Lagrangian. The procedure (see for example \cite{16}) has to be
performed in 2 steps in order to avoid several subsidiary conditions: all
the information for dynamics must be obtained from the same variational
procedure. The starting effective Lagrangian is

\begin{equation}
L_{eff}=\theta f_{\mu \nu }^{\ast }f^{\mu \nu }+\frac{\theta }{2\lambda }%
f^{\mu \nu }\left( \nabla _{\mu }h_{\nu }-\nabla _{\nu }h_{\mu }\right) +%
\frac{\theta }{2\lambda }f^{\ast \mu \nu }\nabla _{\rho }T_{\mu \nu }^{\rho
}+A\overline{\Psi }\left[ \left( \rho ^{\mu }h_{\mu }+m\right) \right] \Psi
+Bh_{\mu }h^{\mu }  \tag{20}
\end{equation}%
where 
\begin{align}
\text{ }\rho _{\mu }& =\left( \mathbf{a}_{c}+\mathbf{b}_{c}\gamma
^{5}\right) \gamma _{\mu }^{c}+\varepsilon _{\mu }^{\alpha \beta \gamma
}\left( \mathbf{c}_{c}\gamma _{\alpha }^{c}\sigma _{\beta \gamma }\right) 
\tag{21} \\
& =\left[ \mathbf{a}_{c}+(\mathbf{b}_{c}+\mathbf{c}_{c})\gamma ^{5}\right]
\gamma _{\mu }^{c}  \notag
\end{align}%
that corresponds to the decomposition of a general vector element of the Lie
algebra of SU(2, 2). Let us notice that if $\mathbf{b}_{c}+\mathbf{c}_{c}=0$%
, the pseudo vectorial part of $\rho _{\mu }$ is eliminated. This fact is
directly related, due to the variational procedure of the effective
Lagrangian (20), with the generalized Hodge-de Rham decomposition that we
have considered before, that is 
\begin{align}
A\overline{\Psi }\rho ^{\mu }\Psi & =\overline{\Psi }\left( \mathbf{a}%
^{c}\gamma _{c}^{\mu }\right) \Psi +\overline{\Psi }\left( \left( \mathbf{b}%
^{c}+\mathbf{c}^{c}\right) \gamma ^{5}\gamma _{c}^{\mu }\right) \Psi 
\tag{22} \\
& =B\underset{h^{\mu }}{\underbrace{\left[ \nabla ^{\mu }\Omega +\varepsilon
^{\mu \beta \gamma \delta }\nabla _{\beta }A_{\gamma \delta }+\gamma
^{5}b^{\mu }+\left( P^{\mu }-eA^{\mu }\right) \right] }}\,.  \notag
\end{align}%
Then, following the standard procedure (Berestetsky et al. \cite{16}) after
deriving the equations of motion and the related constraints, the effective
Lagrangian takes the form%
\begin{equation}
L_{eff}=L_{1}+L_{2}+L_{3}  \tag{23}
\end{equation}%
being 
\begin{align}
L_{1}& =\left( \theta +\lambda \right) f_{\mu \nu }^{\ast }f^{\mu \nu
}\rightarrow (theta\text{ }term)  \tag{24} \\
L_{2}& =A\overline{\Psi }\left[ \left( \rho ^{\mu }\left( P_{\mu }-eA_{\mu
}+\gamma ^{5}b_{\mu }+\nabla _{\mu }\Omega +\varepsilon _{\mu }^{\beta
\gamma \delta }\nabla _{\beta }A_{\gamma \delta }\right) +m\right) \right]
\Psi \rightarrow (Dirac-like\text{ }term)  \tag{25} \\
L_{3}& =A^{2}\overline{\Psi }\rho _{\mu }\Psi \overline{\Psi }\rho ^{\mu
}\Psi \rightarrow (4-fermion-term)  \tag{26}
\end{align}%
It is important to note also that all dependence on coefficient values are
charged on the respective parameters in order to avoid the unboundedness
problem for the Lagrangian (eg: $\theta ,A,etc.).$

\subsection{Energy-momentum tensor and cosmological term}

It is worth noticing that the mass term in the Dirac equation (15) contains
the GR curvature scalar plus the cosmological term $\lambda=(1-d)$. In the
analysis already made in \cite{20}, the mass is a constant, then it is
naturally included into the Dirac equation and then into the energy-momentum
tensor. Also here, the gravitational part of the Lagrangian (containing the
curvature) has been avoided. We can write the effective energy-momentum
tensor derived from the effective Lagrangian density $L_{eff}$, as

\begin{align*}
T_{\rho \sigma }& \propto 4\left( \theta +\lambda \right) \left[ f_{\alpha
\rho }^{\ast }f_{\text{ }\sigma }^{\alpha }-g_{\rho \sigma }\frac{f_{\mu \nu
}^{\ast }f^{\mu \nu }}{4}\right] -A\overline{\Psi }\left[ \left( g_{\rho
\sigma }\rho ^{\mu }h_{\mu }-\left( \rho _{\rho }h_{\sigma }+h_{\rho }\rho
_{\sigma }\right) \pm \frac{\left( \overset{\circ }{R}+\lambda d\right)
g_{\rho \sigma }\mp \overset{\circ }{R}_{\rho \sigma }}{\sqrt{\overset{\circ 
}{R}+\lambda d}}\right) \right] \Psi \\
& -2A^{2}\left[ \frac{g_{\rho \sigma }}{2}\overline{\Psi }\rho _{\mu }\Psi 
\overline{\Psi }\rho ^{\mu }\Psi -\overline{\Psi }\rho _{\rho }\Psi 
\overline{\Psi }\rho _{\sigma }\Psi \right] \,.
\end{align*}%
Using the Dirac equation and rearranging the 4-fermion term, the above
tensor can be rewritten in order to identify the effective contribution to
the cosmological term from the fermion sector. We obtain%
\begin{align*}
T_{\rho \sigma }& \propto 4\left( \theta +\lambda \right) \left[ f_{\alpha
\rho }^{\ast }f_{\text{ }\sigma }^{\alpha }-g_{\rho \sigma }\frac{f_{\mu \nu
}^{\ast }f^{\mu \nu }}{4}\right] -A\overline{\Psi }\left[ \left( -\left(
\rho _{\rho }h_{\sigma }+h_{\rho }\rho _{\sigma }\right) \mp \frac{\overset{%
\circ }{R}_{\rho \sigma }}{\sqrt{\overset{\circ }{R}+\lambda d}}\right) %
\right] \Psi \\
& +A^{2}g_{\rho \sigma }\overline{\Psi }\rho _{\mu }\Psi \overline{\Psi }%
\rho ^{\mu }\Psi \,.
\end{align*}%
As firstly pointed out by Eddington \cite{21}, the mass term is directly
related to the curvature and implied by the Mach principle. Here, we want to
stress that also fermion interactions can contribute to the cosmological
term and then can take part to the cosmic dynamics as a sort of dark energy
contribution \cite{capolupo}. Notice that from the above expression, the
pure fermionic contribution to the cosmological constant, due to the
4-fermion interaction is%
\begin{equation*}
\Lambda _{f}\equiv \kappa \rho _{\Lambda _{f}}=+\kappa A^{2}g_{\rho \sigma }%
\overline{\Psi }\rho _{\mu }\Psi \overline{\Psi }\rho ^{\mu }\Psi
\end{equation*}%
(where the units of the constant are $\left[ \kappa \right] =m_{Pl}^{-2}).$
Considering the possibility of quark condensates, it was conjectured [24]
that a nonzero vacuum expectation value of the 4-fermion term arises from a
spontaneous breaking of the global chiral symmetry by the $\left\langle 
\overline{q}q\right\rangle $ condensate, which sets the energy scale of the
condensation to the $QCD$ scale of the running strong-interaction coupling, $%
\Lambda _{QCD}.$ To see this, the Shifman-Vainshtein-Zakharov (SVZ)
approximation can be effectively used [25] given the following result 
\begin{equation*}
\left\langle 0\left\vert \Lambda _{f}\right\vert 0\right\rangle =\frac{%
16\kappa A^{2}}{9}\left( \mathbf{a}_{c}\mathbf{a}^{c}-\left( \mathbf{b}_{c}+%
\mathbf{c}_{c}\right) \left( \mathbf{b}^{c}+\mathbf{c}^{c}\right) \right)
\left\langle 0\left\vert \overline{\Psi }\Psi \right\vert 0\right\rangle
^{2}\,.
\end{equation*}%
Here, the contribution corresponding to the axial/axial vector channel is
identically zero (only \textbf{A-A}\ and \textbf{V-V}\ expectation values
give contributions to the cosmological constant (see the explicit channel
computations below) and the arbitrary constants $\mathbf{a}_{c},\mathbf{b}%
_{c}$ and $\mathbf{c}_{c}$ can be defined accordingly. It is also useful to
consider that, from the above formula, all the parameters can be fixed from
the corresponding experimental data. The traces for vector and axial-vector
channels explicitly are%
\begin{align*}
\mathrm{Tr}\left[ \left( \widehat{k^{\prime }}+M\left( k^{\prime }\right)
\right) \gamma _{\mu }\left( \widehat{k}+M\left( k\right) \right) I\right] &
=4\left( k^{\prime }M\left( k\right) +kM\left( k^{\prime }\right) \right)
_{\mu }, \\
\mathrm{Tr}\left[ \left( \widehat{p_{2}}+m\right) \gamma _{\mu }\left( 
\widehat{p_{1}}+m\right) \gamma _{5}\right] & =0, \\
\mathrm{Tr}\left[ \left( \widehat{p_{2}}+m\right) \gamma _{\mu }\left( 
\widehat{p_{1}}+m\right) \gamma _{\nu }\right] & =4\left[ m^{2}g_{\mu \nu
}+\left( p_{2\mu }p_{1\nu }-\left( p_{1}p_{2}\right) g_{\mu \nu }\right)
+p_{2\mu }p_{1\nu }\right] , \\
\mathrm{Tr}\left[ \left( \widehat{p_{2}}+m\right) \gamma _{\mu }\left( 
\widehat{p_{1}}+m\right) \gamma _{\nu }\gamma _{5}\right] & =4\varepsilon
_{\alpha \mu \beta \nu }p_{2\alpha }p_{1\beta }, \\
\mathrm{Tr}\left[ \left( \widehat{p_{2}}+m\right) \gamma _{\mu }\left( 
\widehat{p_{1}}+m\right) \sigma _{\lambda \rho }\right] & =4
\end{align*}%
and then the above result $\left\langle 0\left\vert \Lambda _{f}\right\vert
0\right\rangle $ is explicitly recovered.

\section{Discussion and conclusions}

Let us now analyze each term of the Lagrangian (23).\ There is a possible
screening between $\theta $ plus $\lambda $ Lagrangian terms. A similar
relation between $\theta $ and $\lambda $ has been conjectured in Ref. \cite%
{19}.

There are no Holst term and FMT term, in contrast with other theories
involving gravitation in canonical formulation (on the possibility of
getting Holst-like terms in f(R) theories was analized in\cite{28}and\cite%
{29}). The vector-vector and the axial-axial terms are also in the FMT
Lagrangian but the term that we have here is not constrained by any
extra-parameter as the Barbero-Immirzi one. It is clear that the term
proportional to the axial-vector coupling is not present into the model
discussed here due to the fundamental geometrical structure of our
construction. Here we have%
\begin{equation*}
A\overline{\Psi }\rho ^{\mu }\Psi =\overline{\Psi }\left( \mathbf{a}%
_{c}\gamma _{\mu }^{c}\right) \Psi +\overline{\Psi }\left( \left( \mathbf{b}%
_{c}+\mathbf{c}_{c}\right) \gamma ^{5}\gamma _{\mu }^{c}\right) \Psi \equiv 
\mathbf{V+A}
\end{equation*}%
where the sum of axial and vector terms appears. Immediately we see that the
4-fermion interaction $\overline{\Psi }\rho _{\mu }\Psi \overline{\Psi }\rho
^{\mu }\Psi $ geometrically only picks \textbf{V-V} and \textbf{A-A}
interactions, if we suppose coefficients real (in particular $\mathbf{b}_{c}$
and $\mathbf{c}_{c}$). This important fact has been experimentally probed as
pointed out in \cite{22}. In that paper, the effective 4-fermion interaction
was focused on the case of neutrinos endowed with non-standard interactions.
These are a natural outcome of many neutrino mass models \cite{13} and can
be of two types: flavour-changing (FC) and non-universal (NU). As it is well
known, see-saw-type models leads to a non-trivial structure of the lepton
mixing matrix characterizing the charged and neutral current weak
interactions. This leads to gauge induced non standard interactions which
may violate lepton flavor and CP even with massless neutrinos.
Alternatively, non-standard neutrino interactions may also arise in models
where neutrino masses are \textquotedblleft calculable\textquotedblright\
from radiative corrections.

Finally, in some supersymmetric unified models, the strength of non-standard
neutrino interactions may be a calculable renormalization effect. How
sizable are non-standard interactions will be a model-dependent issue. In
some models, non-standard interaction strengths are too small to be relevant
for neutrino propagation, because they are suppressed by some large scale
and/or restricted by limits on neutrino masses. However, this could not be
the case, and there are interesting models where moderate strength
non-standard interactions remain in the limit of light (or even massless)
neutrinos. Such a fact may occur even in the context of fully unified models
like SO(10). Non-standard interactions may, in principle, affect neutrino
propagation properties in matter as well as the detection cross sections.
Thus their existence can modify the solar neutrino signal observed by the
experiments.

There appears, at effective level, a $\theta $ (parity violating) term that
is not present in other formulations as the standard Einstein-Cartan (see
for example the discussion in \cite{20}) and loop-quantum gravity inspired 
\cite{18,19}. For quarks, a non-zero vacuum expectation value of the
4-fermion term arises from a spontaneous breaking of the global chiral
symmetry by the $\left\langle \overline{q}q\right\rangle $ condensate, which
sets the energy scale of the condensation to the QCD scale of the running
strong-interaction coupling, $\Lambda _{QCD}$. Quark condensates are
associated to the color degree of freedom, and characterize the confined
phase of quark matter and constitute, together with gluon condensates, the
QCD vacuum. For leptons, which do not interact strongly and are not
subjected to confinement, less is known about the form and scale of
condensation. Is important to note that in the study of cosmological
constant one should bear in mind that vector torsion gives the contribution
to conformal anomaly (see, for instance, \cite{30})which may give
qualitatively same effect to cosmology (anomaly-driven inflation or LCDM DE)
as pure cosmological constant.

Solving the question about the interplay between gravity models with
torsion, space-time dimensionality and 4-fermion interaction is still
complicated, although some claims in the recent literature point out the
contrary \cite{23}.

The basic points under discussion are: dimensional compactification in
space-times with torsion, the origin of 4-fermion interaction and the
specific extra-dimensional models that have to be considered. From the
viewpoint of the model presented here, the main question to be addressed is,
in an space-time with dimensionality $>$ 4, that the torsion dual will not
be a vector but a higher rank tensor field. However, the total antisymmetry
of the torsion in such a case, simplifies any physical analysis. In a
forthcoming paper, we will discuss the possible experimental tests of the
unified scheme presented here.

\section{Acknowledgements}

DJCL\ is very grateful to the people of the Bogoliubov Laboratory of
Theoretical Physics \ (BLTP)\ and JINR\ Directorate by they hospitality and
financial support, and also to Professors J.W.F. Valle and F.J.

Escrihuela for bring me important references on the subject.

\end{document}